\documentclass[12pt]{iopart}

\usepackage{iopams}  
\usepackage{graphicx}   
\usepackage{array}
\usepackage[dvipdfx,cmyk]{xcolor}     
\usepackage[caption=false]{subfig}  
\usepackage[colorlinks,linkcolor=red,citecolor=brown,urlcolor=blue]{hyperref}
\usepackage{dsfont,bm}
\usepackage{cite}

\newcommand{\be}{\begin{equation}}
\newcommand{\ee}{\end{equation}}
\newcommand{\ben}{\begin{eqnarray}}
\newcommand{\een}{\end{eqnarray}}
\newcommand{\bes}{\begin{subequations}}
	\newcommand{\ees}{\end{subequations}}
\newcommand{\bF}{\begin{figure}}
	\newcommand{\eF}{\end{figure}}

\newcommand{\avg}[1]{\langle {#1} \rangle}

\begin{document}

\title[Ensemble inequivalence and quasi-stationary states in long-range random networks]{Ensemble inequivalence and absence of quasi-stationary states \\in long-range random networks}

\author{L Chakhmakhchyan$^1$, T N Teles$^2$, S Ruffo$^3$}

\address{$^1$Centre for Quantum Information and Communication, Ecole Polytechnique de Bruxelles,
	CP 165, Universit\'{e} libre de Bruxelles, 1050 Brussels, Belgium}

\address{$^2$Departamento de Ci\^{e}ncias Exatas e Sociais Aplicadas, Universidade Federal de Ci\^{e}ncias da Saude de Porto Alegre, CEP 90050-170, Porto Alegre, RS, Brazil}

\address{$^3$SISSA, Via Bonomea 265, INFN and ISC-CNR, I-34136 Trieste, Italy}
\ead{levon.chakhmakhchyan@ulb.ac.be}
\vspace{10pt}

\begin{abstract}
Ensemble inequivalence has been previously displayed only for long-range interacting systems with non-extensive energy. In order to perform the thermodynamic limit, such systems require an unphysical, so-called, Kac rescaling of the coupling constant. We here study models defined on long-range random networks, which avoid such a rescaling. The proposed models have an extensive energy, which is however non-additive. For such long-range random networks, pairs of sites are coupled with a probability decaying with the distance $r$ as $1/r^\delta$. In one dimension and with $0 \leq \delta <1$, the surface energy scales linearly with the network size, while for $\delta >1$ it is $O(1)$. By performing numerical simulations, we show that a negative specific heat region is present in the microcanonical ensemble of a Blume-Capel model, in correspondence with a first-order phase transition in the canonical one. This proves that ensemble inequivalence is a consequence of {\it non-additivity} rather than {\it non-extensivity}. Moreover, since a mean-field coupling is absent in such networks, relaxation to equilibrium takes place on an intensive time scale and quasi-stationary states are absent.
\end{abstract}



\section{Introduction}

A wide range of problems in physics concerns Long-Range Interacting systems (LRIs), 
systems embedded in $d$ dimensions and characterized by a pairwise potential decaying at large distances as $1/r^\delta$ ($0 \leq \delta\leq d$). Examples include self-gravitating systems, non-neutral plasmas, geophysical flows, etc.~\cite{rev, rev1}. The energy of LRIs is non-additive, which determines {\it ensemble inequivalence}~\cite{BC}. This entails unusual thermodynamic features, such as negative specific heat, within the microcanonical ensemble~\cite{grav,grav1,fel,BC}. LRIs possess  intriguing collective macroscopic phenomena as well. Namely, such systems may become trapped in out-of-equilibrium quasi-stationary states, whose lifetime diverges with system size~\cite{rev, rev1, qss}. This behavior is caused by the presence of mean-field couplings in the kinetic theory description of LRIs, which implies an underlying Vlasov equation~\cite{Nicholson}.

The statistical and dynamical properties of LRIs have been mainly studied for systems without disorder and defined on regular lattices, although a few studies have been devoted to the random-field Ising model and to spin glasses~\cite{Nishimori}. Meanwhile, the understanding of long-range coupling effects on complex structures remains a tantalising problem, which has attracted considerable attention recently. Specifically, ensemble inequivalence has been analysed for Bethe lattices~\cite{potts}, and quasi-stationary states have been found for diluted Erd\"{o}s-R\'{e}nyi random graphs~\cite{Ciani}. Additionally, unusual convexity properties of thermodynamic functions for models defined on random networks have been discussed from a mathematical perspective~\cite{Radin}. Finally, a possible alternative origin of ensemble inequivalence for LRIs on random graphs was proposed~\cite{Garlaschelli}, and the presence of phase transitions for several models defined on scale-free, small-world~\cite{herrero, herrero1} and other random networks was reviewed~\cite{Dorogovtsev}.

Long-Range Random Networks (LRRNs) were introduced independently by Leuzzi et al.~\cite{parisi} and Daqing et al.~\cite{sern}. An important aspect of such networks is the inclusion of the Euclidean distance separating the nodes. The probability of linking 
two nodes at a given distance $r$ decays as $p(r,\delta) \sim 1/r^{\delta}$, which mimics long-range couplings. Interestingly,  LRRNs are capable of modelling real networks, whose link lengths are characterised by a power-law distribution such as Internet, global airlines, power grids \cite{sern}. The statistical properties of LRRNs were analyzed from several perspectives, including, in particular, the emergence of phase transitions in their structural (such as mean topological distance and clustering coefficient) and percolation properties for some critical values of the control parameter $\delta$~\cite{epl,epl1}. Furthermore, using canonical Monte-Carlo simulations, ferromagnetic transitions in random networks ($\delta=0$), as well as mean-field-to-Ising universality crossovers with respect to the value of the $\delta$ exponent were recently revealed ~\cite{manhatt,herrero}. Moreover, LRRNs found their application in efforts to optimise traffic dynamics and to design optimal transportation networks~\cite{transport,epl2}. Importantly, spin models defined on LRRNs (where bonds are Gaussian distributed with a zero mean and a variance obeying a power-law distribution with the distance) are also easier to simulate than their fully connected versions and allow one to probe the spin-glass phase beyond the mean-field regime~\cite{parisi}.

In this paper, we address the problem of ensemble inequivalence in LRRNs. Namely, we begin with the study of the Blume-Capel (BC) model defined on a LRRN.  Despite the presence of long-range interactions, the energy of the BC model defined on LRRNs is naturally {\it extensive} being otherwise {\it non-additive}. Consequently, one does not require here any artificial Kac rescaling of the coupling~\cite{kac}. Additionally, this system exhibits all necessary features for displaying inequivalence between microcanonical and canonical ensembles, such as the presence of a tricritical point. These important properties allow us to demonstrate that ensemble inequivalence is a consequence of the {\it non-extensivity} of the system rather than of its {\it non-additivity}. We then address dynamical effects on LRRNs by considering the Hamiltonian Mean Field model~\cite{rev,rev1,qss}. We show that quasi-stationary states are absent, which suggests that these states are a consequence of the {\it mean-field} nature of the interactions.

\section{The Blume-Capel model defined on long-range random networks}

We define the BC model on a complex network $\mathcal{C}$ in terms of the following Hamiltonian
\begin{equation}\label{1}
\mathcal{H}_\mathrm{BC}=-\frac{J}{2 K}\sum_{i,j=1}^N A_{ij}^\mathcal{C}S_iS_j+\Delta\sum_{i=1}^NS_i^2,
\end{equation}
where $N$ spin-1 variables $S_i=-1, 0, 1$ occupy the nodes of a network defined by means of its adjacency
matrix $A_{ij}^\mathcal{C}$. The strength of ferromagnetic couplings is given by $J>0$, while  $\Delta>0$ controls the energy difference between magnetic and non-magnetic states.

Let us concisely describe the iterative construction of the adjacency matrix of LRRNs~\cite{sern, parisi},  following the prescription given in \cite{lrnn, epl, epl1}.
Starting from a regular one-dimensional lattice with $N$ sites (the construction is easily 
generalized to $d$ dimensions), we assign a fixed integer number
$k_{\mathrm{max}}$ to all nodes, which defines their maximal degree. At each iteration of the process
we add a link of unit strength to a pair of randomly chosen sites at a distance $r$ with a probability $p(r, \delta)=c r^{-\delta}$, only if both have an available link. Here, $c$ is determined
from the normalization condition $\int_{1}^{N}p(r, \delta)dr=1$, yielding, for large $N$, $c=\delta-1$ ($\delta>1$) and $c=(1-\delta)N^{\delta-1}$ ($0\leq\delta<1$)~\cite{lrnn}. We perform up to $10^3 N$ iterations\footnote{The number of iterations necessary to generate an Erdos-Renyi random network is of order $O(N^2)$, proportional
	to the number of pairs of nodes. Here, iterations scale with $N$ because the maximal number of links per node is fixed to $k_{\mathrm{max}}$.
} and, at the end of the process, we 
remove multiple links. This results into the construction of an irregular directed complex network. However, one has to fulfil the symmetry of the adjacency matrix $A_{ij}^\mathcal{C}$. Therefore, having established a link between nodes $i\rightarrow j$, we add
a symmetric one, $j\rightarrow i$. The described construction of the adjacency matrix $A_{ij}^\mathcal{C}$, given the above definition of $p(r, \delta)$, is well-defined for any practical realization of LRRNs with a finite number $N$ of network nodes~\cite{lrnn}. Our numerical simulations of the statistical (internal energy, magnetization) and dynamical properties of LRRNs, presented in the following sections, show a convergence in the thermodynamic limit $N\to \infty$. The latter observation is further confirmed by previous studies of order-disorder and percolation transitions, as well as of structural features of such networks~\cite{epl, epl1, epl2, manhatt}. We therefore strongly believe that the model is well-defined in the thermodynamic limit.

In general, systems involving long-range interactions require the Kac rescaling factor $K \sim N$ in equation~(\ref{1}),
which allows one to keep the energy of the system extensive~\cite{kac}. A remarkable feature of the BC model defined on LRRNs 
is that, despite the presence of long-range couplings, it does not require the Kac rescaling.
More precisely, as the maximal degree of the network, $k_{\mathrm{max}}$, is fixed, the energy of the system
scales linearly with $N$, i.e. it is {\it extensive}. For a configuration with all spins aligned,
the coupling energy reads $-J/(2 K)\sum_{i,j=1}^N A_{ij}^\mathcal{C}S_iS_j \approx -JN \avg{k}/(2 K)$,
where $\avg{k}$ is the average degree of the nodes ($\avg{k}\approx k_\mathrm{max}$~\cite{lrnn}).
Thus, in order to make the energy independent of $k_{\mathrm{max}}$, we simply set $K\equiv k_{\mathrm{max}}
\approx \avg{k}$.

A crucial parameter that defines the properties of our model is the exponent $\delta$ of the coupling probability 
$p(r, \delta)$. We are interested here in the range $0 \leq \delta < 1$, which leads to non-additive energies. Namely, we expect that the total energy $E_\mathrm{tot}$ of the system is not equal, in the large $N$ limit, to the sum of the energies of its two
components $E_1$ and $E_2$, consisting of $N/2$ spins each. In order to illustrate this property, we consider a configuration with all spins aligned and we set $\Delta=0$. In this case $E_\mathrm{tot}= -JN/2$ and the energies $E_1$ and $E_2$ are given by the number of links within each component of the lattice~\cite{lrnn}. It is then easy to check that for $\delta > 1$ and $N \gg 1$, the energies of the first and second components are $E_1=E_2=-JN/4$ and thus the surface energy $E_\mathrm{tot}-(E_1+E_2)$ is of $O(1)$. 
For $0 \leq \delta < 1$, we find $E_1=E_2=-JN2^{\delta-3}$, yielding $E_\mathrm{tot}-(E_1+E_2)=-JN(1-2^{\delta-1})/2$. Therefore, in this range of $\delta$, the surface energy scales linearly with system size $N$ and remains comparable with the bulk energies $E_1$ and $E_2$, even in the large $N$ limit.

\section{Microcanonical and canonical ensembles}

We study numerically the equilibrium properties of model (\ref{1}) in both the microcanonical and the canonical ensembles.
For the microcanonical simulations, we use Creutz's Monte Carlo algorithm~\cite{creutz}, which introduces an
auxiliary degree of freedom, termed ``demon", with non-negative initial energy $E_\mathrm{D}$. Taking the system 
with energy $E$, determined by the initial microscopic configuration, one attempts to make a single spin flip. If the energy of the system is lowered by this flip, it is accepted and the excess energy is 
given to the demon. If the attempted flip results in an increase in energy, the necessary energy 
is taken from the demon and the flip is carried out. If the demon does not have enough energy, the flip is rejected and 
another one is attempted. One can show that, under this dynamics, and to leading order in the system size, the energy 
distribution of the demon takes the form $P(E_D)\sim e^{-E_D/T_{\mu}}$. Thus, measuring this distribution, yields the 
microcanonical temperature $T_{\mu}$, which corresponds to the energy $E$. Typically, we perform $10^5-10^6$ sweeps to obtain convergence to equilibrium. The canonical ensemble is simulated using the standard Metropolis algorithm~\cite{metro}.

\begin{figure}[ht]
	\begin{center}
	\includegraphics[width=6.5cm]{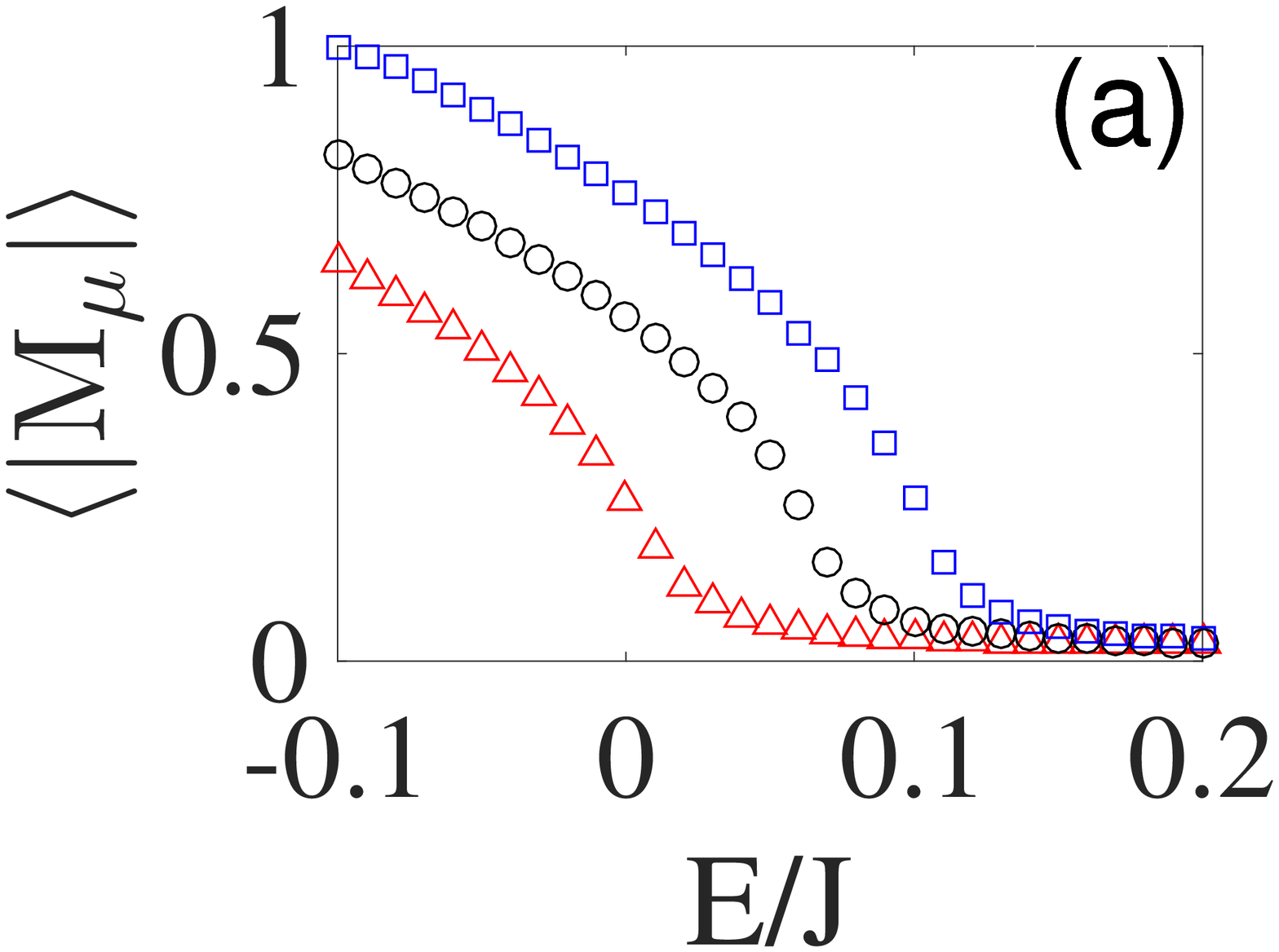}
	\includegraphics[width=6.5cm]{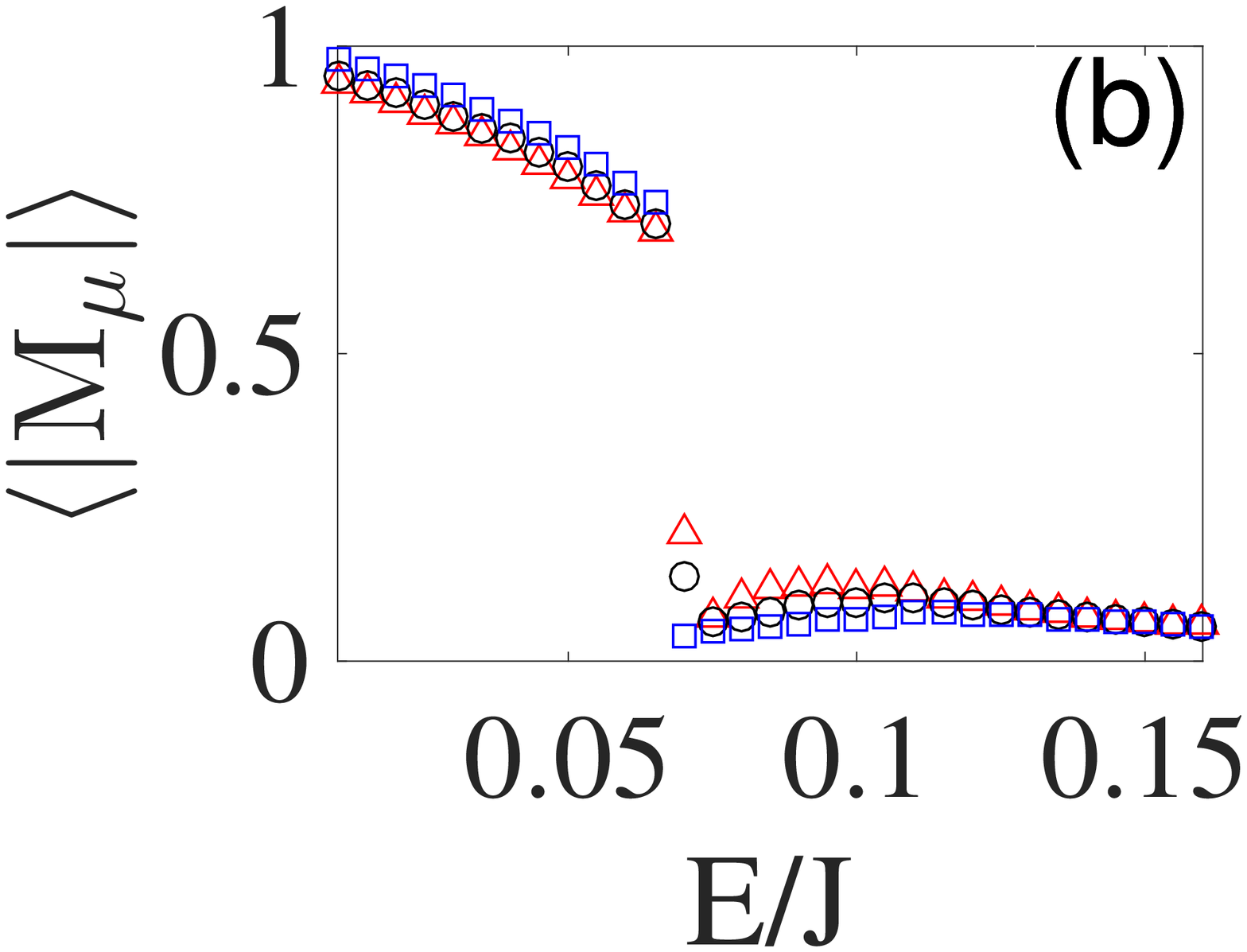}
	\caption {Modulus of the microcanonical magnetization $\langle|M_{\mu}|\rangle$ of the BC model on a LRRN versus the energy $E$ in the region of (a) second-order phase transitions ($\Delta=0.2$ (red triangles), $\Delta=0.3$ (black circles) and $\Delta=0.4$ (blue squares));  (b) first-order phase transitions ($\Delta=0.485$ (red triangles), $\Delta=0.49$ (black circles) and $\Delta=0.5$ (blue squares)). For both cases $k_\mathrm{max}=4$ and $\delta=0.1$. The number of spins is fixed to $N=500$, but we have checked that the behavior of the magnetization remains stable with respect to a further increase of $N$. \label{mag}}
	\end{center}
\end{figure}

In figure~\ref{mag}(a) we plot the modulus of the microcanonical magnetization $\avg{|M_\mu|}$, $M_{\mu}=1/N\sum_{i=1}^NS_i$, versus the energy $E$\footnote{We have computed the modulus of the magnetization rather than the magnetization itself in order to avoid sign inversions near the transition energy caused by finite $N$ effects. This results into a slightly larger magnetization, but still of the order $1/\sqrt{N}$, in the paramagnetic phase.}. This figure shows how $\avg{|M_\mu|}$, for small values of $\Delta$ (panel (a)),  gradually vanishes when increasing energy, showing a second-order phase transition, while it jumps to zero for larger values of $\Delta$ (panel  (b)), corresponding to a first order phase transition. We then expect that model (\ref{1}) has a {\it tricritical point} in the range $\Delta=[0.4,0.485]$, in a similar way as the “mean-field” BC model~\cite{BC}. 

\begin{figure}[h!]
	\begin{center}
		\includegraphics[width=8cm]{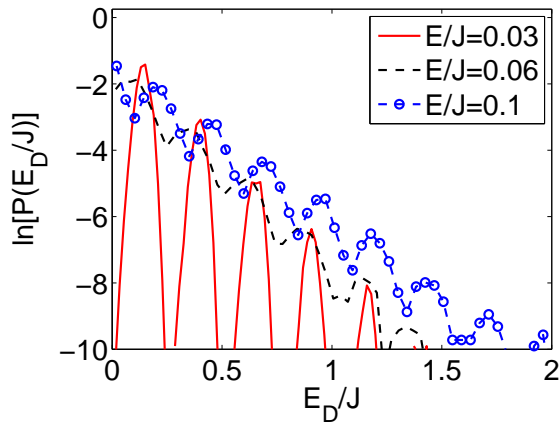}
		\caption {Logarithm of the probability distribution, $P(E_\mathrm{D}/J)$, of the rescaled demon energy $E_\mathrm{D}/J$ for different initial configurations of fixed energy $E$. Here $N=900$, $k_\mathrm{\mathrm{max}}=4$, $\delta=0.1$ and $\Delta/J=0.495$. \label{hist}}
	\end{center}
\end{figure}

Next, we address the issue of defining microcanonical temperature $T_{\mu}$ in terms of the probability distribution of the demon's energy. As figure~\ref{hist} shows, the corresponding histogram consists of several peaks, their envelope obeying an exponential decay. Moreover, each of these peaks can be well fitted by a Gaussian function, centered around the points corresponding to energies $\Delta E$ exchanged between the system and the demon as a result of a spin flip (the sharpness of the peaks becomes less pronounced when increasing the energy of the initial configuration). Specifically, these values of $\Delta E$ depend on the degree of the node at which the flipped spin is nested, on the configuration of the spins linked to it and on the initial and flipped values of the central spin itself. Importantly, due to the inherent structure of the LRRN, the degree of every node is random, determined in terms of the adjacency matrix $A_{ij}^\mathcal{C}$. It is precisely this randomness of the coupling matrix, present in the expression of $\Delta E$, that is responsible for the peaked structure of $P(E_D/J)$. We have indeed checked that 
the probability distribution of the BC model on a regular one-dimensional lattice with nearest-neighbor couplings does not possess the aforementioned peaked structure, but exhibits a purely exponential decay. Moreover, the usual exponential behavior of $P(E_\mathrm{D}/J)$ is 
recovered also when the adjacency matrix $A_{ij}^\mathcal{C}$ describes a regular network, where all the nodes have equal degree. We also note that for the case of the BC model on LRRNs with $\Delta=0$, $P(E_\mathrm{D}/J)$ consists of extremely narrow peaks centered at the exchange energies $\Delta E$, each of which is a multiple of $J/k_\mathrm{max}$. The envelope of $P(E_D/J)$ is a perfect exponential, which, analogously to the BC model on regular lattices or in the mean-field case, defines the microcanonical temperature $T_{\mu}$. This is also confirmed by the equivalence of canonical and microcanonical temperatures in the low and high-energy regions, as we discuss below.

\begin{figure}[ht]
	\begin{center}
		\includegraphics[width=6.5cm]{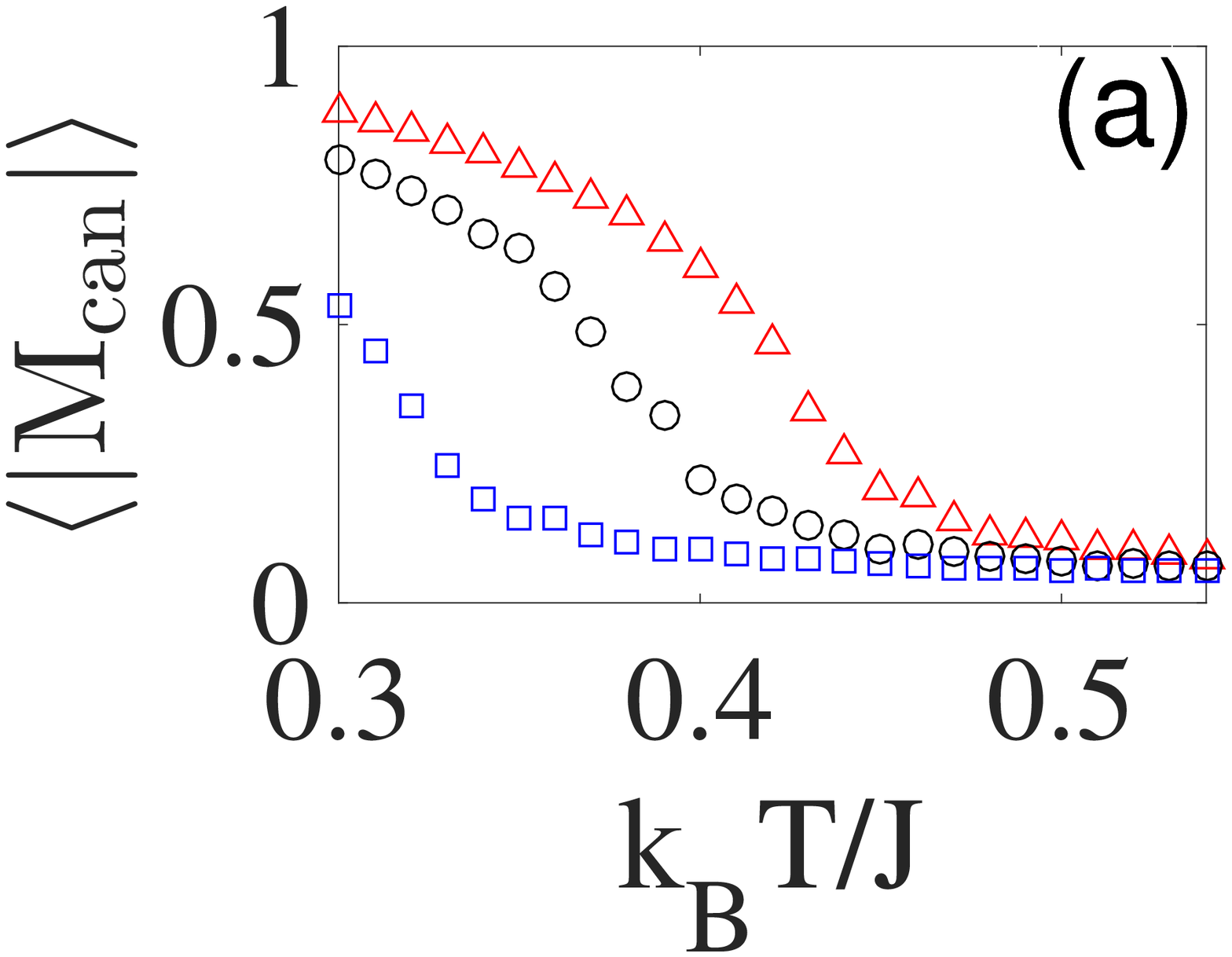}
		\includegraphics[width=6.5cm]{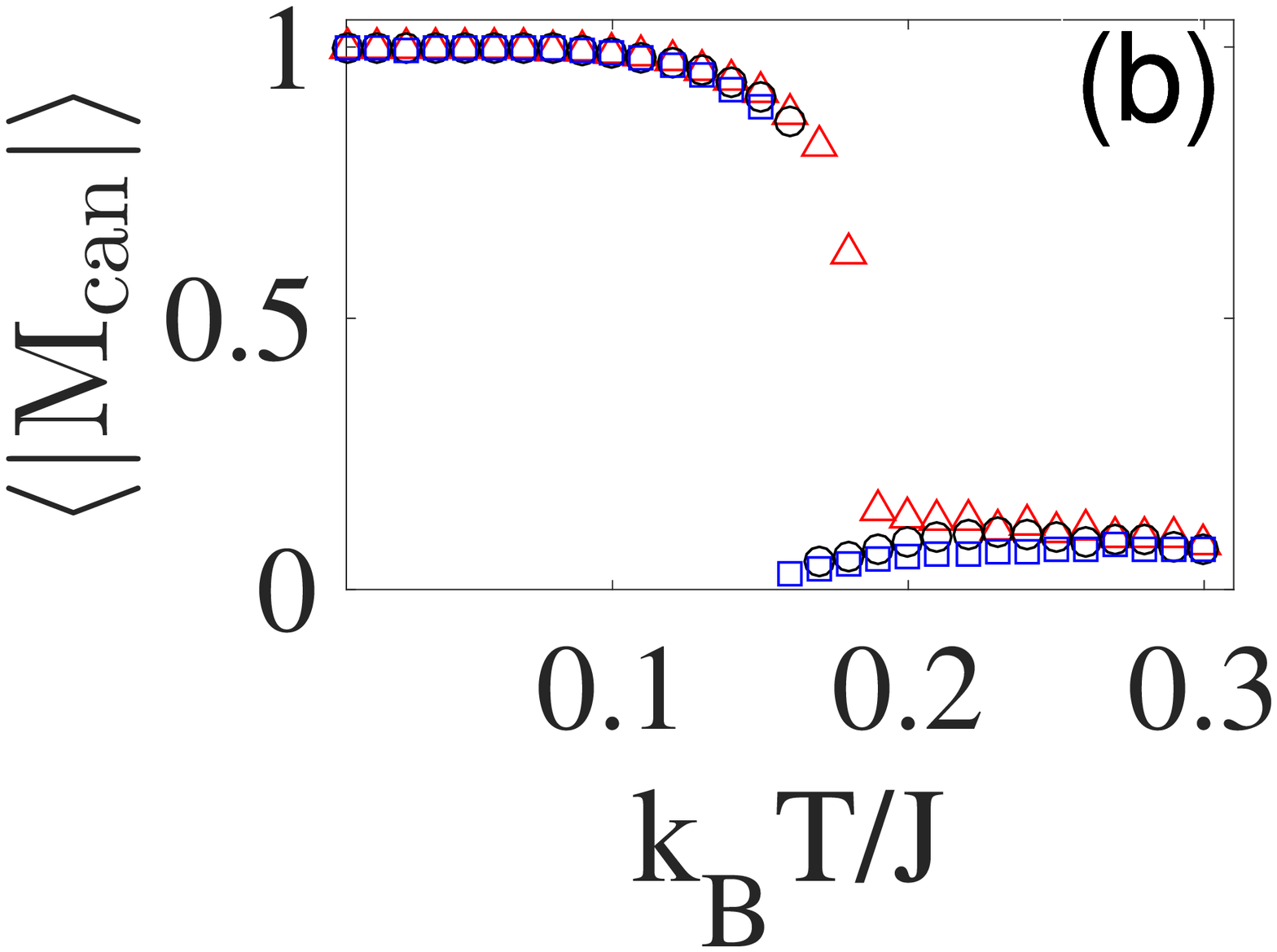}
		\caption {Modulus of the canonical magnetization $\langle|M_{\mathrm{can}}|\rangle$  of the BC model on a LRRN versus the temperature $T$ in the region of (a) second-order phase transitions ($\Delta=0.2$ (red triangles), $\Delta=0.3$ (black circles) and $\Delta=0.4$ (blue squares)); (b) first-order phase transitions ($\Delta=0.485$ (red triangles), $\Delta=0.49$ (black circles) and $\Delta=0.5$ (blue squares)). For both cases $k_\mathrm{max}=4$, $\delta=0.1$ and $N=500$.  \label{magmetr}}
	\end{center}
\end{figure}

In figure~\ref{magmetr} we show the dependence of magnetization $\langle|M_\mathrm{can}|\rangle$ versus temperature $T$
in the canonical ensemble, confirming the presence of first- and second-order phase transitions and of a tricritical point.

The phase diagram of the BC model on LRRNs in the $(\Delta/J,T/J)$ plane should not differ much from the one of the mean-field model 
\cite{BC}, displaying a line of second-order phase transitions at small $\Delta/J$ ending in a tricritical point. At larger values
of $\Delta/J$ the transition is first order and ends at zero temperature at a limiting value $\Delta_c/J$. This value can be deduced 
by the following argument: at zero temperature, the energy at transition in the two phases are equal. In the paramagnetic 
phase, the energy $E_\mathrm{par}$ vanishes ($E_\mathrm{par}=0$), while in the ordered phase the majority of spins is parallel and the corresponding energy can be approximated as $E_\mathrm{ord}\approx -JN\langle k \rangle/(2k_\mathrm{max})+\Delta N$. As at the phase transition $E_\mathrm{ord}=E_\mathrm{par}$, we obtain the estimate $\Delta_c\approx J \langle k \rangle/(2k_\mathrm{max})\approx J/2$. 

\section{Ensemble inequivalence}
\begin{figure}[ht]
	\begin{center}
		\includegraphics[width=6.5cm]{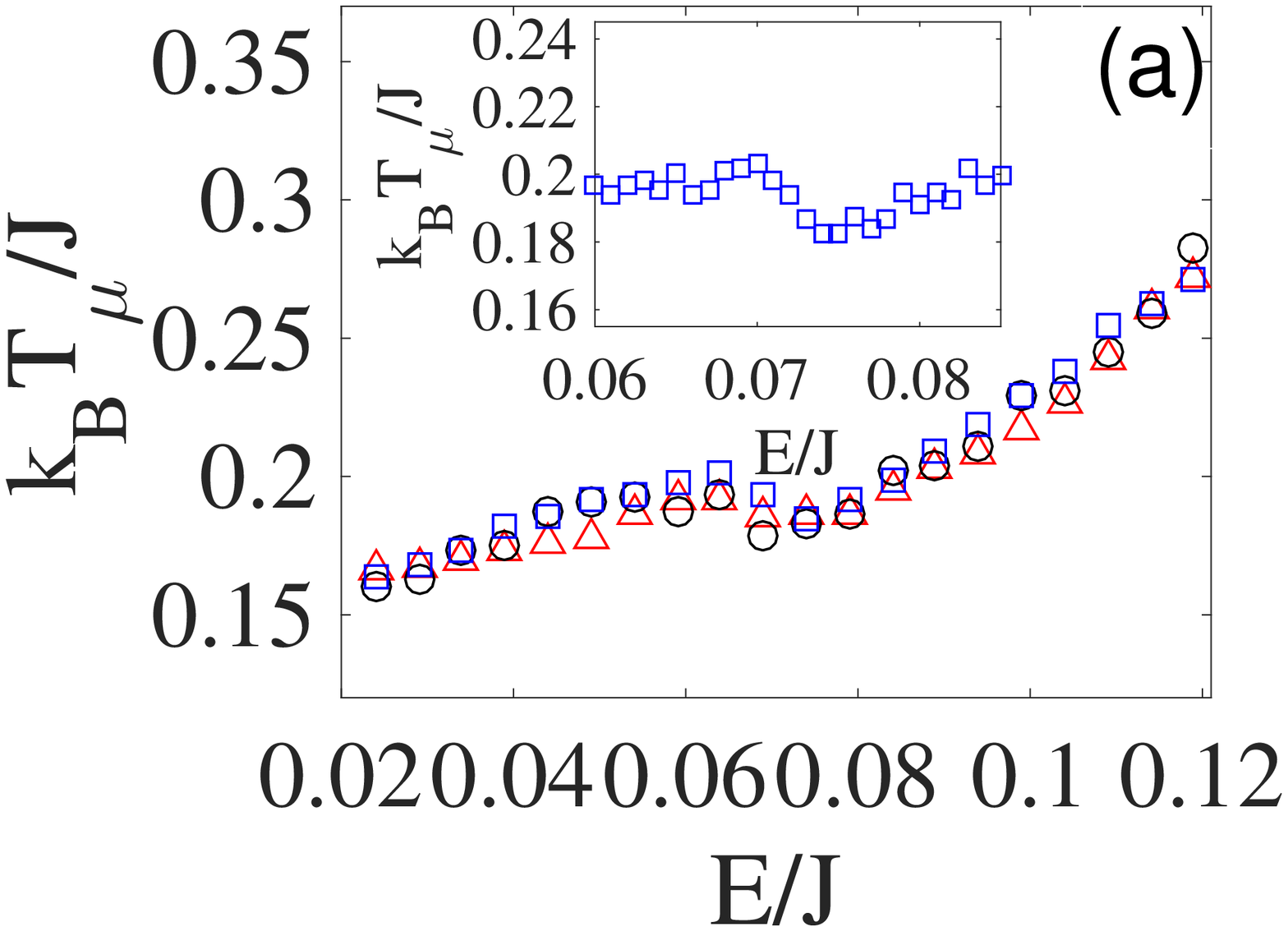}
		\includegraphics[width=6.5cm]{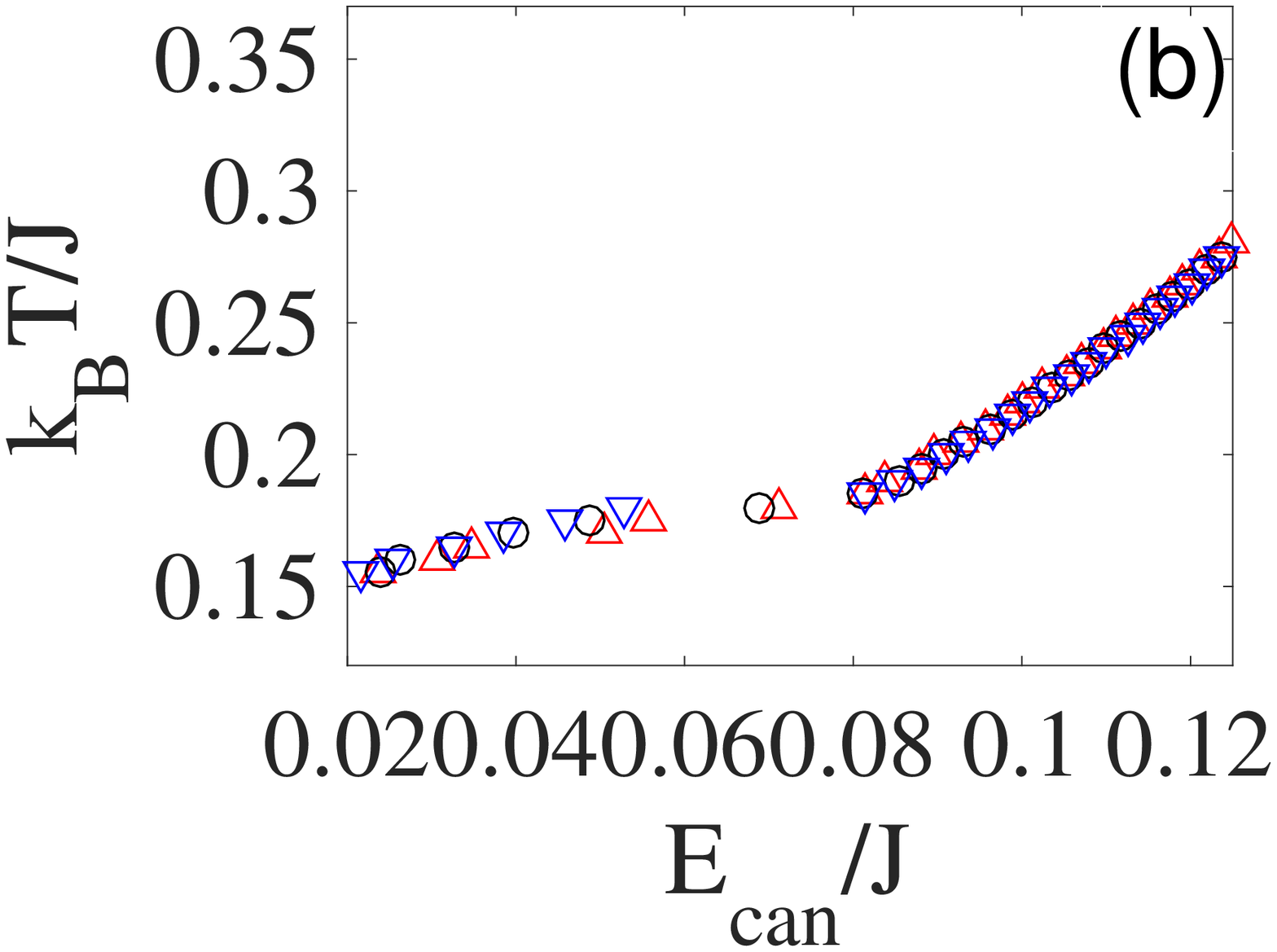}
		\caption {The microcanonical (a) and canonical (b) caloric curves of the BC model on a LRRN for different numbers of spins: $N=300$ (red triangles), $N=600$ (black circles), $N=900$ (blue squares). For both cases $k_\mathrm{max}=4$, $\delta=0.1$ and $\Delta/J =0.485$. The inset in (a) shows the details in the region of ensemble inequivalence (the bumpy behavior of the curve is due to the intrinsic randomness of LRRNs). \label{calor}}
	\end{center}
\end{figure}

\begin{figure}[ht]
	\begin{center}
		\includegraphics[width=8cm]{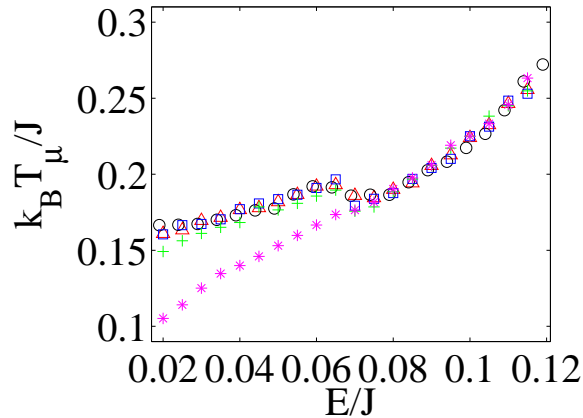}
		\caption {The microcanonical caloric curves of the BC model on a LRRN for different values of the exponent $\delta$: $\delta=0$ (red triangles), $\delta=0.1$ (black circles),  $\delta=0.5$ (blue squares),  $\delta=0.9$ (green crosses), $\delta=1.3$ (magenta asterisks). Here, $\Delta=0.485$ and $k_\mathrm{max}=4$. The number of spins is fixed to $N=600$, but we have checked that the behavior of the magnetization remains stable with respect to a further increase of $N$. \label{deltas}} \end{center}
\end{figure}

The main purpose of simulating the properties of the BC model on LRRNs in canonical and microcanonical ensembles is to test their equivalence. We achieve this by constructing the caloric curves of the system in both ensembles. Namely, we consider the dependence of the microcanonical temperature $T_\mu$ on energy and of the canonical mean energy $E_{\mathrm{can}}$ on temperature in the vicinity of the first-order phase transition point. As figure~\ref{calor} shows, in the high-energy region the two ensembles are equivalent. However, when lowering the energy, in the vicinity of the first-order phase transition point, the microcanonical caloric curve exhibits a negative specific heat, not captured by the canonical ensemble (note that we expect that in the large $N$ limit, the almost flat region in figure~\ref{calor}(b) near the transition temperature  becomes exactly flat, showing the jump in energy corresponding to the latent heat from the paramagnetic to the ferromagnetic phase). 
The different behavior of the caloric curve in the two ensembles thus reveals the inequivalence of the canonical and microcanonical ensembles. To further illustrate this inequivalence, we plot in figure~\ref{deltas} caloric curves for several values of the coupling probability exponent $\delta$. As this figure shows, the negative specific heat behavior is indeed present for $0\leq \delta<1$, confirming inequivalence of ensembles in this region of $\delta$. Moreover, as expected, it vanishes when $\delta>1$, due to the recovered additivity of the model, resulting in the equivalence of ensembles. We emphasize once again that the system energy is non-additive, but is extensive and we therefore deal here with a novel class of systems which exhibits ensemble inequivalence.

\section{Absence of quasi-stationary states in long-range random networks}

\begin{figure}[ht]
	\begin{center}
		\includegraphics[width=8cm]{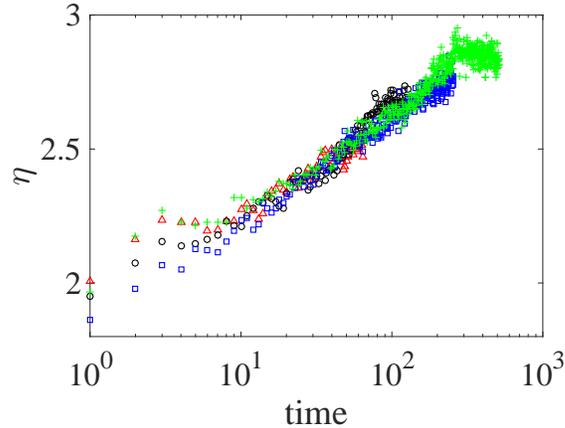}
		\caption {Time evolution of the kurtosis $\eta$ in the HMF model defined on LRRN (\ref{2}) for different system sizes: 
			$N=64$ (red squares), $N=128$ (black circles), $N=256$ (blue squares), $N=512$ (green crosses). Here $\delta=0.1$, $p_\mathrm{max}=1.5$ 
			and the total energy $E = 0.3$ (in order to obtain smooth curves, several realizations of the random initial condition and of the 
			LRRN are performed). 
			\label{kurt}}	\end{center}
\end{figure}

In order to test the dynamical behavior on LRRNs, we consider the Hamiltonian Mean Field (HMF) model~\cite{qss}
\begin{equation}\label{2}
\mathcal{H}_\mathrm{HMF}=\sum_{i=1}^N \frac{p_i^2}{2}-\frac{J}{2 k_\mathrm{max}}\sum_{i,j=1}^N A_{ij}^\mathcal{C}\cos (\theta_i - \theta_j),
\end{equation}
where $A_{ij}^\mathcal{C}$ is the adjacency matrix of the LRRN, $\theta_i \in [-\pi,\pi]$ is the angular degree of freedom of node $i$ and $p_i$ is the conjugated momentum (one does not require Kac rescaling here). In the mean-field case this model exhibits
Quasi-Stationary States (QSS) that relax to equilibrium on a time scale diverging with system size $N$~\cite{rev, rev1}. We check
the presence of such states by considering an initial state of zero magnetization and a ``water-bag" momentum distribution, 
choosing random angles $\theta_i \in [-\pi,\pi]$ and a random momentum $p_i\in [-p_{\textrm{max}}, p_{\textrm{max}} ]$ 
($p_\textrm{max}$ is adjusted, in order to keep the system at zero magnetization). 
We expect to observe a relaxation to a Gaussian distribution in momentum. We detect the relaxation time by monitoring the evolution of the kurtosis
of the momentum distribution $\eta=\avg{p^4}/\avg{p^2}^2$, which takes the value $\eta_{\mathrm{eq}}=3$ at equilibrium. 
As figure~\ref{kurt} shows, the relaxation to equilibrium takes place on an intensive time scale, i.e. the relaxation time does
not increase with $N$. This behavior proves that there are no QSS in the HMF model defined on LRRNs. 
Therefore, we conclude that the appearance of such states occurs due to the mean-field nature of interactions. 

\section{Conclusion}
In this paper we have studied the statistical and dynamical properties of models defined on Long-Range Random Networks (LRRNs). 
As we have shown, the energy of the Blume-Capel (BC) model is extensive but non-additive in a certain region of the decay rate
of the coupling probability, $0\leq \delta <1$, and does not require the Kac rescaling prescription to perform the thermodynamic
limit. This model has certain similarities with the mean-field BC model, but it does not admit an analytical solution. We have thus 
simulated the model numerically in both the microcanonical and the canonical ensemble using Creutz and Metropolis Monte Carlo algorithm, respectively. 
Both algorithms confirmed the presence of first- and second-order phase transitions. Moreover, in the region $0\leq \delta <1$ and in the vicinity of the 
first-order phase transition we have observed a negative specific heat in the microcanonical ensemble, a clear signature of ensemble
inequivalence. For larger values of $\delta$ we fall into a short-range regime, due to the recovered additivity of the model, and the
negative specific heat region disappears. Therefore, the proposed model belongs to a novel class of {\it non-additive} but {\it extensive} 
long-range interacting systems, whose existence clarifies that non-additivity is the true origin of ensemble inequivalence. Furthermore, 
in order to explore dynamical effects on LRRNs, we have considered the Hamiltonian Mean Field (HMF) model and shown that its relaxation 
time to equilibrium does not increase with system size, thus showing the absence of Quasi-Stationary States (QSS). We thus achieved a second 
important conclusion of this paper, that QSS appear due to the {\it mean-field} nature of the interactions and not as a consequence of
the presence of long-range couplings.

\ack
L.C. acknowledges support from the H2020-FETPROACT-2014 Grant QUCHIP (Quantum
Simulation on a Photonic Chip; grant agreement no. 641039).

\end{document}